# OPEN IT TOOLS FOR AN INQUIRY-BASED PHYSICS EDUCATION

Martin Šechný, martin.sechny@shenk.sk

*Shorter version of this paper has been accepted at the 24th conference of Slovak physicists (2019). http://kf.elf.stuba.sk/~24konferencia/*

## INTRODUCTION

In case of inquiry-based education, the priority is the pupil's activity, developing his/her practical and research skills. We can use the knowledge and skills obtained by the pupil from other subjects. Applied informatics into physics seems to be a suitable application of digital literacy. Open IT tools are the ones of the available IT tools that bring several benefits to promote freedom of inquiry.

An overview of open IT tools for physics education includes microprocessor kits with a variety of analog and digital electronic components, a large number of application software, simulation software, programming languages, libraries and development environments, web tools, virtual laboratories, data catalogs, hypertext documents and other auxiliary resources. Such open IT tools together with suitable open educational resources will be applicable in physics, mathematics, informatics and other subjects.

## INQUIRY-BASED PHYSICS EDUCATION

Physics is a natural science subject that forms the natural or technical thinking of the pupil. The key to the success of teaching physics is the motivation of the pupil. Inquiry-based learning simulates the work of a scientist in exploring, discovering, researching. Inquiry-based education fits best on natural science subjects, including physics, because natural sciences are the discovering sciences. Inquiry-based education is an activation method where the pupil is active and the teacher accompanies him. Inquiry-based learning is derived initially from constructivism, self-creation of a problem/task solution and also from constructionism, self-formation of knowledge (learning by making). The pupil is naturally curious.

Inquiry-based education can be realized at several levels of inquiry [1]:

- Interactive demonstration
- Exploration (verification, confirmation)
- Directed inquiry (instructed experimentation)
- Guided inquiry (teacher assigns the task)
- Open inquiry (pupil assigns the task)

The tasks should focus on different levels of cognitive functions: manipulation with real objects, mental representations, abstract concepts.

## APPLIED INFORMATICS INTO PHYSICS

Informatics as a school subject should be a mix of all major fields of study: computer engineering, computer science, data science. Physics, together with mathematics, gives informatics a science base (computer science) and a technical base (computer engineering). Applied informatics in theoretical and experimental physics is the engine of research and development. Applied informatics (digital skills, programming and data science) in physics can be a new trend in didactics of physics that could increase pupils' motivation and practical use of knowledge and skills.

Open tools are suitable in the inquiry-based education. Openness, expressed in an appropriate public license, gives the user the right to use the tool for any purpose – for example, to explore something with the tool, to explore how the tool works, to modify the tool. This is the freedom of inquiry. The low price of the open tool is also a significant factor.

Teamwork is an important part of inquiry-based education. That's why we also need collaborative IT tools.

## MICROPROCESSOR KITS

We usually call a group of science and technology subjects as STEM (science, technology, engineering, mathematics).[1] Various practical STEM activities in a classroom use hands-on approach. [2] Microprocessor kits are suitable tools for the interleaving of physics and informatics. These kits consist of a small printed circuit board with a microprocessor/microcontroller and a set of various electronic components.

The best-known open source hardware kits are: BBC micro:bit, Arduino, Raspberry Pi, with a galore of all similar products.

BBC micro:bit[2] (fig. 1)[3] is an inexpensive 8/16/32-bit ARM-based microcontroller with a set of sensors and LEDs, especially suitable for primary schools. The pupils can attach various sensors by electric wires, make portable or wearable devices.

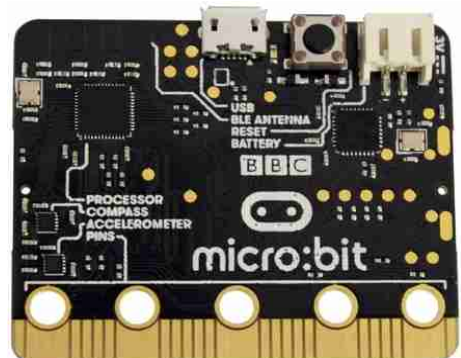

Fig. 1. BBC micro:bit

---

1. https://en.wikipedia.org/wiki/Science,_technology,_engineering,_and_mathematics
2. https://en.wikipedia.org/wiki/Micro_Bit, https://microbit.org
3. https://cdn11.bigcommerce.com/s-am5zt8xfow/images/stencil/500x659/products/1203/2968/apivkgqpf__10493.1548550465.jpg

Arduino[4] (fig. 2)[5] is an 8/16/32-bit minicomputer with analog and digital I/O, mostly based on Atmel AVR microcontroller. Arduino kit contains a set of external electronic components, extensions or add-ons.

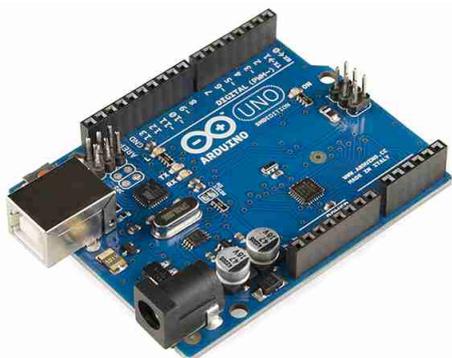

Fig. 2. Arduino

Raspberry Pi[6] (fig. 3)[7] is a 64-bit computer based on Broadcom BCM microprocessor, with a full operating system (GNU/Linux).

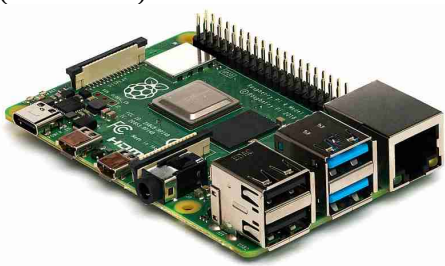

Fig. 3. Raspberry Pi

Arduino and Raspberry Pi are suitable mostly for secondary schools. Some microprocessor kits could be interconnected to form more complex configurations, for example the Raspberry Pi with an extension board together with analog sensors or actuators (as parts of the Arduino kit).

**OPEN SOURCE SOFTWARE, FREE SOFTWARE**

Public-funded schools should preferably use open source software / free software for education and for operation. Open source software is easily portable to different operating systems and different hardware.

GNU/Linux[8] is a famous open source operating system. This software offers several attractive graphical user interfaces, a command line (bash) that is a versatile efficient tool for processing text and numeric data and for scripting. The GNU[9] project unites many developers and forms an extensive collection of free software.

CERN/Fermilab Scientific Linux[10] is a special operating system distribution designed for scientists, but actually discontinued. There are some similar distributions focused on science.

Android[11], the Linux-based most popular operating system for smartphones and tablets, is easily accessible to pupils and teachers. There are many Android applications using computer sensors. We can find interesting guides for inquiry-based learning with Android devices in the classroom on the web. [3]

Algodoo[12] is an application for simple physical animations. Algodoo is a freeware, but not an open source software.

Tracker[13] makes visualization and motion analysis of video (fig. 4)[14]. This is free and open source software.

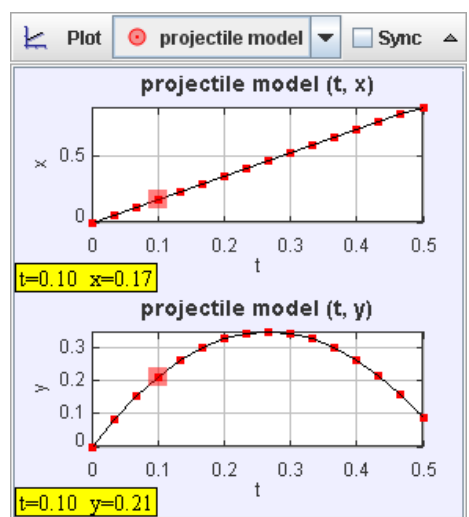

Fig. 4. Tracker: plot

EJS[15] (Easy Java/JavaScript Simulations) helps non-programmers create interactive simulations in Java or JavaScript, mainly for teaching or learning purposes. EJS has free non-commercial license and commercial license.

Popular programming environments for younger pupils: Scratch[16], MIT App Inventor[17]. Programming languages suitable for education and science: Python, PHP, JavaScript, C/C++, Java and others. Languages for data storing, processing and presentation: SQL, HTML/XHTML and CSS, XML, JSON.

Geogebra[18] is easy-to-use mathematical application for pupils. Geogebra has free non-commercial license.

Gnuplot[19] is a useful tool for visualizing numeric data. Gnuplot is a freeware and open source software with some restrictions (Gnuplot is not a part of GNU).

---

4  https://en.wikipedia.org/wiki/Arduino,
   https://www.arduino.cc
5  https://sk.wikipedia.org/wiki/S%C3%BAbor:Arduino_Uno_-_R3.jpg
6  https://en.wikipedia.org/wiki/Raspberry_Pi,
   https://www.raspberrypi.org
7  https://en.wikipedia.org/wiki/File:Raspberry_Pi_4_Model_B_-_Side.jpg
8  https://en.wikipedia.org/wiki/Linux,
   https://distrowatch.com
9  https://www.gnu.org
   https://en.wikipedia.org/wiki/GNU
10 https://www.scientificlinux.org
11 https://en.wikipedia.org/wiki/Android_(operating_system),
   https://www.android.com
12 http://www.algodoo.com
13 https://physlets.org/tracker/
14 https://physlets.org/tracker/help_images/plot_double.gif
15 http://fem.um.es/Ejs/
16 https://scratch.mit.edu
17 https://appinventor.mit.edu
18 https://www.geogebra.org
19 http://www.gnuplot.info

Powerful mathematical software, programming languages or libraries: Scilab[20], Octave[21], Yacas[22], Maxima[23], SciPy[24], R[25], SageMath[26], GDL[27] etc. KNIME[28] is a data analytics, reporting and integration platform.

LibreCAD[29] and QCAD[30] are 2D technical charts editors. KiCad[31], ngspice[32], logisim[33] are applications for electronics.

Finally common applications: Mozilla Firefox[34] browser, LibreOffice[35] suite including Calc spreadsheet which is usually sufficient to analyze data at school.

## WEB TOOLS

We need collaborative tools in inquiry-based education [4], preferably web tools. Etherpad[36] is an easy open source web-based collaborative real-time text editor. We can be inspired by programmers who use complex collaborative tools: Slack[37], Git[38].

Didactically processed educational content is provided by interactive web games, interactive web-based educational applications, open courses (MOOC).[39]

Wikipedia[40], free online encyclopedia, is probably the first choice when looking for expert information. Wikidata[41] acts as central storage for the structured data of projects including Wikipedia and others. Wikidata can be used also as a data source for school topics.

Wolfram Alpha[42] is a computational knowledge engine, very interesting for pupils and teachers. Wolfram Alpha is free for use on the web. As an application, it has commercial license, however it is included in Raspberry Pi software for free non-commercial use (Wolfram Language & Mathematica)[43].

---

20  https://www.scilab.org
21  https://www.gnu.org/software/octave/
22  http://www.yacas.org
23  http://maxima.sourceforge.net
24  https://www.scipy.org
25  https://www.r-project.org/about.html
26  https://www.sagemath.org
27  https://github.com/gnudatalanguage/gdl
28  https://www.knime.com
29  https://librecad.org
30  https://qcad.org
31  http://kicad-pcb.org
32  http://ngspice.sourceforge.net
33  http://www.cburch.com/logisim/
34  https://www.mozilla.org/sk/firefox/
35  https://www.libreoffice.org
36  https://etherpad.org
37  https://slack.com
38  https://git-scm.com
39  https://en.wikipedia.org/wiki/Massive_open_online_course, https://en.wikipedia.org/wiki/List_of_MOOCs
40  https://www.wikipedia.org
41  https://www.wikidata.org/wiki/Wikidata:Main_Page
42  https://www.wolframalpha.com
43  https://www.raspberrypi.org/documentation/usage/mathematica/README.md

## VIRTUAL LABORATORIES

If the school does not have a real laboratory, it can use a virtual laboratory. Watching video of some experiment is the simplest form. ASPIRE Lab[44] provides a simulated laboratory, similar to real scientific simulations. Computer virtualization allows you to run a software environment with the required parameters for computer-based experiments. Controlled remote experiment is a typical scientific method, which is also useful in the education. Virtual observatory is a special virtual laboratory, mainly for astronomy and astrophysics, such as EURO-VO project[45]. Some software tips: Stellarium[46] (virtual planetarium) (fig. 5), Celestia[47] (interactive space simulator), Aladin[48] (interactive space atlas with data processing and visualization).

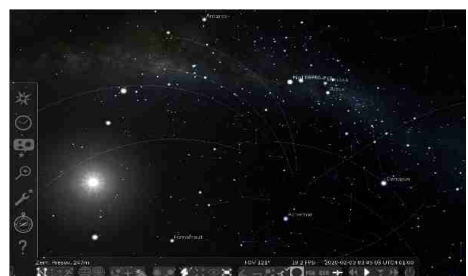

Fig. 5. Stellarium

## CONCLUSIONS

Inquiry-based learning increases pupil's motivation through attractive observations and experiments, and develops practical and research skills. Applied informatics in physics is probably an effective way of applying digital literacy to physical thinking. Open IT tools bring benefits and promote freedom of inquiry.

---

44  E. g. https://aspire.eecs.berkeley.edu
45  http://www.euro-vo.org
46  https://stellarium.org
47  https://celestia.space
48  https://aladin.u-strasbg.fr